\documentstyle[12pt]{article}
\newcounter{mycount}

\newcommand{\be}{\begin{eqnarray}}
\newcommand{\ee}{\end{eqnarray}}
\newcommand{\bfl}{\begin{flushleft}}
\newcommand{\efl}{\end{flushleft}}

\newcommand\noi{\noindent}

\begin{document}

\bibliographystyle{nphys}

\centerline {CLIFFORD ALGEBRA OF TWO-FORMS,}
\centerline {CONFORMAL STRUCTURES, AND FIELD EQUATIONS}
\vspace* {-25 mm}
\begin{flushleft} ITP 92-41 \\
September 1992
\end{flushleft}

\vskip 0.9in
\centerline{Ingemar Bengtsson }

\vskip 2cm
\centerline{\bf Abstract}
\vskip 2 cm

I review the equivalence between duality operators on two-forms and conformal
 structures in four dimensions, from a Clifford algebra point of view (due to
Urbantke
 and Harnett). I also review an application, which leads to a set of
"neighbours" of
Einstein's equations. An attempt to formulate reality conditions for the
"neighbours"
is discussed.
\vskip 6cm
Talk presented at the second Max Born Symposium on "Spinors, Twistors and
Clifford
 Algebras", Wroclaw, September 1992

\vskip 3mm
\vfill\noi
\begin {flushright} Institute of Theoretical  Physics

S-41296 G{\"o}teborg

Sweden
\end {flushright}

\vskip 3mm \noi

\eject

There is a deep theory for how to solve the self-dual Yang-Mills equations

\vspace{3mm}

\begin{equation}*F_{{\alpha}{\beta}i} =
g^{-1/2}g_{{\alpha}{\gamma}}g_{{\beta}{\delta}}
\tilde{F}_i^{{\gamma}{\delta}} = 1/2
g^{-1/2}g_{{\alpha}{\gamma}}g_{{\beta}{\delta}}{\epsilon}^{{\gamma}{\delta}{\mu}{\nu}}F_{{\mu}{\nu}i} = F_{{\alpha}{\beta}i}\end{equation}

\vspace{3mm}
\noindent
where the duality operator is defined with respect to some fixed conformal
structure, i.e. a metric up to a conformal factor (and some useful notation -
the twiddle - has been introduced as well). Some time ago it occurred to
Urbantke (1984) to pose this problem backwards: Given a field strength, with
respect to which conformal structure is it self-dual? There is an elegant
solution to this curious question, and an elegant proof - due to Urbantke and
Harnett (1991) - based on the Clifford algebra of two-forms in four dimensional
spaces. For the moment, let me state the result and then indicate how I want to
use it.
  We need a triplet of two-forms, which is non-degenerate in the sense that it
may serve as a basis in the three-dimensional space of self-dual two-forms. In
particular, the index i ranges from one to three. Then

\vspace{3mm}

\begin{equation}
g_{{\alpha}{\beta}} = - 2/3
{\eta}f_{ijk}F_{{\alpha}{\gamma}i}\tilde{F}^{{\gamma}{\delta}j}F_{{\delta}{\beta}k}
\end{equation}
\noindent
is Urbantke's formula. It gives the metric with respect to which
$F_{{\alpha}{\beta}i}$ is automatically self-dual (the $f_{ijk}$ are the
structure constants of SO(3), and the conformal factor ${\eta}$ is so far
arbitrary).

  Some work by Capovilla, Jacobson and Dell (1989) may be regarded as a more
ambitious version of Urbantke's formula. (See also Plebanski 1977, Capovilla,
Dell, Jacobson and Mason 1991.) We may regard $F_{{\alpha}{\beta}i}$ as the
self-dual part of the Riemann tensor, considered - at the outset - as just an
SO(3) field strength, with no connection to the metric. Then the question
arises whether it is possible to formulate a set of differential equations,
using the SO(3) connection (and the Levi-Civita tensor densities) alone, such
that the above metric becomes Ricci flat. The answer turns out to be yes; more
specifically, the answer is the field equations following from the action

\vspace{3mm}

\begin{equation} S = 1/8 \int {\eta}(Tr{\Omega}^2 - 1/2(Tr{\Omega})^2)
.\end{equation}

\vspace{3mm}
\noindent
where ${\eta}$ is a Lagrange multiplier and

\vspace{3mm}

\begin{equation} {\Omega}_{ij} =
{\epsilon}^{{\alpha}{\beta}{\gamma}{\delta}}F_{{\alpha}{\beta}i}F_{{\gamma}{\delta}j}
\end{equation}

\vspace{3mm}
\noindent
The existence of this action is closely related to Ashtekar's (1987)
formulation of the 3+1 version of Einstein's theory - in fact the CDJ action is
a natural Lagrangian formulation of Ashtekar's variables. The action which
leads to Einstein's equations including a cosmological constant is less
elegant.

  The next question is: What happens if we use the above building blocks to
write an arbitrary action

\vspace{3mm}

\begin{equation}S = \int{\cal L}({\eta}; Tr{\Omega}, Tr{\Omega}^2,
Tr{\Omega}^3) ,\end{equation}

\vspace{3mm}
\noindent
where the only restriction on $\cal L$ is that it has density weight one? (Due
to the characteristic equation for three-by-three matrices, there are only
three independent traces.) The action is certainly generally covariant. Suppose
that we solve the field equations and use Urbantke's formula to define a
metric. Is that reasonable, and relevant for physics? What happens if we change
the structure group from SO(3) to something else?

  Now that you know where I am going, we return to prove Urbantke's formula.
For any four-dimensional vector space $\bf V$, the two-forms give a
six-dimensional vector space $\bf W$, with a natural metric

\vspace{3mm}

\begin{equation}({\Sigma}_1,{\Sigma}_2) = 1/2
{\epsilon}^{{\alpha}{\beta}{\gamma}{\delta}}{\Sigma}_{1{\alpha}{\beta}}{\Sigma}_{2{\gamma}{\delta}} .\end{equation}

\vspace{3mm}
\noindent
There is a corresponding Clifford map to the space of endomorphisms on $\bf V
\oplus \bf V^*$:

\vspace{3mm}

\begin{equation}{\gamma}({\Sigma}) = 2
\left(
\begin{array}{llll}
0 & \tilde{{\Sigma}}^{{\alpha}{\beta}}\\
{\Sigma}_{{\alpha}{\beta}} & 0
\end{array}
\right)
;\hspace{3mm}{\gamma}({\Sigma})^2 = - ({\Sigma}, {\Sigma})\bf{1}.\end{equation}

\vspace{3mm}
\noindent
We see that the original vector space $\bf V$ now becomes the space of Weyl
spinors for the Clifford algebra of two-forms.

   Now we introduce a metric on $\bf V$, so that we can define the duality
operator *. $\bf W$ then splits into two orthogonal subspaces $\bf W^+$
(self-dual forms) and $\bf W^-$ (anti-self-dual forms). We choose Euclidean
signature, so that ** = 1, and without loss of essential generality we choose
the determinant of the metric to equal one. Then

\vspace{3mm}

\begin{equation}{\gamma}(*{\Sigma}) = {\gamma}(Z){\gamma}({\Sigma}){\gamma}(Z)
\end{equation}

\vspace{3mm}
\noindent
where

\vspace{3mm}

\begin{equation}{\gamma}(Z) =
\left(
\begin{array}{llll}
0 & g^{{\alpha}{\beta}}\\
g_{{\alpha}{\beta}} & 0
\end{array}
\right)
.\end{equation}

\vspace{3mm}
\noindent
Using a well-known property of six-dimensional ${\gamma}$-matrices, and Swedish
indices in $\bf W$, we can find a totally anti-symmetric tensor $Z^{\ddot{u}
\ddot{a}\ddot{o}}$ such that

\vspace{3mm}

\begin{equation}g_{{\alpha}{\beta}} = Z^{\ddot{u}\ddot{a}
\ddot{o}}{\Sigma}_{{\alpha}{\gamma}\ddot{u}}\tilde{{\Sigma}}^{{\gamma}{\delta}}_{\ddot{a}}{\Sigma}_{{\delta}{\beta}\ddot{o}} .\end{equation}

\vspace{3mm}
\noindent
This determines Z uniquely, and we observe that

\vspace{3mm}

\begin{equation}*{\Sigma} = Z{\Sigma}Z \hspace{3mm} \Rightarrow \hspace {3mm}
Z{\Sigma} = {\Sigma}Z \hspace{3mm} ({\Sigma} \in W^+) \hspace{5mm} Z{\Sigma} =
- {\Sigma}Z \hspace{3mm} ({\Sigma} \in W^-) .\end{equation}

\vspace{3mm}
\noindent
We need a little bit more information about Z.

   To prove the result we are after, we will commit the atrocity of choosing a
basis in $\bf W$. First we choose an ON-basis in $\bf V$, and then we set

\vspace{3mm}

\begin{equation}M_i = e_0\wedge e_i \hspace{4mm} N_i = 1/2 f_{ijk}e_j\wedge
e_k; \hspace{5mm} X_i = M_i - N_i \hspace{4mm} Y_i = M_i + N_i.\end{equation}

\vspace{3mm}
\noindent
Clearly, the X's (Y's) form a basis for $\bf W^-$ ($\bf W^+$), and

\vspace{3mm}

\begin{equation}(X_i, X_j) = - {\delta}_{ij} \hspace{5mm} (Y_i, Y_j) =
{\delta}_{ij} . \end{equation}

\vspace{3mm}
\noindent
Looking back on eq. (11), we see that we can set

\vspace{3mm}
\begin{equation}Z = Y_1Y_2Y_3 \end{equation}

\vspace{3mm}
\noindent
- that is to say that Z is the unit volume element of $\bf W^+$. But, since
$\bf W^+$ is three-dimensional, this is all we need. In terms of an arbitrary
basis ${\Sigma}_{{\alpha}{\beta}i}$ on $\bf W^+$, eq. (10) now becomes

\vspace{3mm}
\begin{equation}g_{{\alpha}{\beta}} \propto
{\epsilon}^{ijk}{\Sigma}_{{\alpha}{\gamma}i}\tilde{{\Sigma}}^{{\gamma}{\delta}}_j{\Sigma}_{{\delta}{\beta}k} . \end{equation}

\vspace{3mm}
\noindent
This is Urbantke's formula.

   When the metric on $\bf V$ has neutral signature, the metric on $\bf W^+$
becomes indefinite, but the discussion is similar, while it becomes slightly
more subtle if the metric on $\bf V$ is Lorentzian.

  With this understanding of eq. (2), let us return to the action (5). Our main
result so far (Capovilla 1992, Bengtsson and Peld\'{a}n 1992, Bengtsson 1991,
Peld\'{a}n 1992) is that this action admits a 3+1 decomposition, and that the
resulting formalism is a natural generalization of "Ashtekar's variables" for
gravity. As is well known, the constraint algebra of general relativity
actually singles out the space-time metric by means of its structure functions.
For the SO(3) case, it turns out that - up to some ambiguity concerning the
conformal factor - the "Hamiltonian" metric is precisely the same as
Urbantke's. We refer to the models in this class as "neighbours of Einstein's
equations", since they all have the same number of degrees of freedom. I will
not discuss the case of arbitrary structure groups here.

  There are several holes that have to be filled before we can claim that we
have really been able to generalize Einstein's equations in an unsuspected way.
For the case of Euclidean signatures, we have to show that the field equations
derived from the action (5) ensure that the metric (2) is positive definite,
rather than neutral. This can be done in specific cases. As an example,
consider the action

\vspace{3mm}

\begin{equation}S = 1/8 \int {\eta}(Tr{\Omega}^2 +
{\alpha}(Tr{\Omega})^2).\end{equation}

\vspace{3mm}
\noindent
As is clear from the preceding discussion, there must be some property of the
field equations that ensure that the matrix ${\Omega}_{ij}$ has definite
signature. To see this, choose a gauge such that the matrix becomes diagonal.
Then it is a straightforward exercise to show that the constraint that results
when varying the action with respect to ${\eta}$ implies that the matrix
${\Omega}_{ij}$ has definite signature if and only if

\vspace{3mm}

\begin{equation}{\alpha} \geq - 1/2 . \end{equation}

\vspace{3mm}
\noindent
In particular, ${\alpha} = - 1/2$, which leads to Einstein's equations, is all
right. (I owe this observation to Ted Jacobson.) Although it is not quite clear
what a general statement is, it is clear that, in general, the requirement that
the metric should have Euclidean signature will lead to some restrictions on
the allowed actions.

   A similar discussion can be given for neutral signature, provided that the
definition of the traces in the action is appropriately changed.

   Our understanding of the Lorentzian case is in much worse shape. It is
necessary to show that propagation is causal with respect to the metric that we
have defined. Moreover (since self-dual two-forms are necessarily complex in
this case) the variables in the action are complex valued, and one must show
how to impose restrictions that imply that the metric is real in any solution.
I believe that the latter problem is the crucial one, and that the former
property somehow follows from the latter. It will not come as a surprise if I
state that the conformal structure is real if and only if

\vspace{3mm}

\begin{equation}(F_i,\bar{F}_j) = 0 ,\end{equation}

\vspace{3mm}
\noindent
where the bar denotes complex conjugation. However, this condition is not very
helpful in itself. It is not difficult to write down solutions with real
Lorentzian metrics - a small zoo of real solutions is already known, for
various "neighbours" (generalizations of Schwarzschild, de Sitter, Kasner,
...). On the other hand, there will always be some solutions for which the
metric is not real - also in the Einstein case. The correct formulation of the
problem is presumably to require that the space of real solutions should be
"reasonably" big - of the same order as the space of solutions of Einstein's
equations, say. It seems natural to switch to the Hamiltonian form of the
equations, and to address the problem from an initial data point of view.
Unfortunately, as soon as this is done, one discovers that the reality
properties of the metric can be discussed easily (Ashtekar 1987) if and only if
we deal with the Einstein case - for the more general models contained in the
action (5), the calculations tend to b

  Which is where the matter stands at the moment. It is perhaps appropriate to
add that we have investigated, in a preliminary way, whether the "neighbours"
can be used to explain any property of the real world. The preliminary answer
was not very encouraging, but perhaps the final verdict is not in yet.
Certainly the more difficult case of arbitrary structure groups (Peld\'{a}n
1992), which was not discussed here, should be carefully studied in this
regard.

\vspace{1cm}

{\it Acknowledgements:} I thank Helmuth Urbantke and Ted Jacobson for
explaining things, and the organizers for a nice stay in the castle.

\newpage
\noindent
$\bf REFERENCES$

\vspace{5mm}
\noindent
A. Ashtekar (1987): New Hamiltonian Formulation of General Relativity, Phys.
Rev. $\bf D36$ 1587.

\vspace{3mm}
\noindent
I. Bengtsson (1991): Self-Duality and the Metric in a Family of Neighbours of
Einstein's Equations, J. Math. Phys. $\bf 32$ 3158.

\vspace{3mm}
\noindent
I. Bengtsson and P. Peld\'{a}n (1992): Another "Cosmological" Constant, Int. J.
Mod. Phys. $\bf A7$ 1287.

\vspace{3mm}
\noindent
R. Capovilla (1992): Generally Covariant Gauge Theories, Nucl. Phys. $\bf B373$
233.

\vspace{3mm}
\noindent
R. Capovilla, J. Dell and T. Jacobson (1989): General Relativity without the
Metric, Phys. Rev. Lett. $\bf 63$ 2325.

\vspace{3mm}
\noindent
R. Capovilla, J. Dell, T. Jacobson and L. Mason (1991): Self-dual 2-forms and
Gravity, Class. Quant. Grav. $\bf 8$ 41.

\vspace{3mm}
\noindent
G. Harnett (1991): The Bivector Clifford Algebra, unpublished manuscript.

\vspace{3mm}
\noindent
P. Peld\'{a}n (1992): Ashtekar's Variables for Arbitrary Gauge Group, Phys.
Rev. $\bf D$, to appear.

\vspace{3mm}
\noindent
J. Plebanski (1977): On the Separation of Einsteinian Substructures, J. Math.
Phys. $\bf 18$ 2511.

\vspace{3mm}
\noindent
H. Urbantke (1984): On Integrability Properties of SU(2) Yang-Mills Fields. I.
Infinitesimal Part, J. Math. Phys. $\bf 25$ 2321.
\end{document}